\documentstyle[ltwol,epsfig]{article}

\def\NIMA{{\em Nucl. Instrum. and Methods} A}

\def\PRL#1#2#3{{\it Phys. Rev. Lett. }{\bf #1 }(#2) #3}

\def\PL#1#2#3{{\it Phys. Lett. }{\bf B#1 }(#2) #3}
\def\EPJ#1#2#3{{\it Euro. Phys. J.}
{\bf C#1 }(#2) #3}

\def\be{\begin{equation}}
\def\ee{\end{equation}}
\def\bea{\begin{eqnarray}}
\def\eea{\end{eqnarray}}
\newcommand{\Bs}     {\mathrm{B}_{\mathrm{s}}}
\newcommand{\Bz}     {\mathrm{B}^0}
\newcommand{\B}      {\mathrm{B}}
\newcommand{\dms}    {\Delta m_{\mathrm{s}}}
\newcommand{\dgs}    {\Delta \Gamma_{\mathrm{s}}}
\newcommand{\Ds}     {\mathrm{D}_{\mathrm{s}}}
\newcommand{\Bd}     {\mathrm{B}_{\mathrm{d}}}
\newcommand{\Bp}     {\mathrm{B}^+}
\newcommand{\dmd}    {\Delta m_{\mathrm{d}}}
\newcommand{\dq}     {\mathrm{d}}
\newcommand{\sq}     {\mathrm{s}}

\newcommand{\tq}     {\mathrm{t}}
\newcommand{\bq}     {\mathrm{b}}
\newcommand{\psin}   {\mathrm{ps}^{-1}}
\newcommand{\ps}     {\mathrm{ps}}
\newcommand{\CL}     {\mathrm{C.L.}}

\newcommand{\amp}    {{\mathcal{A}}}
\newcommand{\siga}   {\sigma_{\mathcal{A}} }

\newcommand{\gev}    {\mathrm{GeV}}
\newcommand{\mev}    {\mathrm{MeV}}
\newcommand{\fs}     {f_{\mathrm{s}}}

\newcommand{\bzb}    {\overline{\mathrm{B}^0}}
\newcommand{\bz}     {{\mathrm{B}^0}}

\def\Aleph{The ALEPH Coll., }
\def\Delphi{The DELPHI Coll., }
\def\L3{The L3 Coll., }
\def\Opal{The OPAL Coll., }
\def\cdf{The CDF Coll., }
\def\sld{The SLD Coll., }

\def\etal{ et al.}

\begin{document}

\boldmath
\title{$\Bs$ PHYSICS AT LEP, SLD, AND CDF: $\dms$ AND $\dgs$}
\unboldmath

\author{GA\"{E}LLE BOIX}

\address{CERN EP-Division, 1211 Geneva 23, Switzerland\\
        E-mail: gaelle.boix@cern.ch}


\twocolumn[\maketitle\abstracts{ The current status of the experimental
knowledge of $\Bs$ meson physics is reviewed. Results from LEP and CDF on the
width difference $\dgs$ are presented, the corresponding average
is found to be in good agreement with the present theoretical estimation.
The $\Bs$ oscillations have not yet been resolved, despite the progress
recently achieved by SLD and ALEPH. The world combination, including
results from the LEP experiments, SLD and CDF,
is presented,
together with the expected and observed lower limit on the $\Bs$ oscillation
frequency. A tantalizing hint of an oscillation is observed around
$\dms\sim17\,\psin$, near future
results could increase the significance of this hint.
}]

\section{Introduction}
\label{sec:intro}
%The $\Upsilon(4S)$ B factories era has began, but still very interesting
%work is carried out with the only $\Bs$ data samples available before Tevatron
%and LHC start collecting data. 
While $\Bz/\Bp$ physics will soon become a monopoly of the new asymmetric
B factories, interesting results are still coming from the existing data
samples containing $\Bs$ mesons.
The LEP experiments have collected around
160K $\Bs$ decays each (equivalent to $\sim300 \,\,\Ds\ell$ candidates reconstructed),
while SLD collected a factor 10 less statistics. In a  different environment,
CDF recorded around 150K $\Bs\to\ell$ decays with roughly $600\,\,\Ds\ell$ 
candidates reconstructed. All these data, collected up to 1998, are diluted 
with nine times more other b-hadrons and are still partially under analysis due to 
the experimental  difficulties involved.

The $\Bs$ mesons are particularly interesting because of their 
particle-antiparticle oscillations. The oscillation frequency, proportional to
the mass difference between the mass eigenstates, $\dms$, is related to the CKM matrix through 
$\dms\sim|V_{\tq\bq}V_{\tq\sq}|$, and combined with the $\Bd$ mass difference,
is related to the CP violation description in the Standard Model:
$\dmd/\dms\sim\sqrt{(1-\rho)^2+\eta^2}$. Physics beyond the Standard Model
could be revealed by a measured value of $\dms$ significantly larger than predicted.

The mass difference $\dms$ is experimentally accessible through 
two complementary methods: $i)$ direct searches for $\Bs$ oscillations 
and $ii)$  measurement of $\dgs$, the width difference
between the two mass eigenstates, as the ratio $\dgs/\dms$ is computable
on the lattice. The $\Bs$ mass eigenstates have a defined CP parity,
%if CP violation in mixing is neglected.
if CP is conserved in mixing.

\boldmath
\section{The $\dgs$ Measurement}\unboldmath
\label{sec:dgs}

The width difference, defined as 
\mbox{$\dgs\equiv\Gamma_{\sq}^{\mathrm{short}}-\Gamma_{\sq}^{\mathrm{long}}$},
($\Gamma_{\sq}^{\mathrm{short}}=\Gamma_{\sq}^{\mathrm{light}}=\Gamma_{\sq}^{\mathrm{even}}$),
has been studied by the four LEP experiments and CDF. 
Experimental information on $\dgs$ can be extracted by studying the proper time
distribution of data samples enriched in $\Bs$ mesons. An alternative method based
on the measurement of the branching ratio 
$\Bs\to\mathrm{D}_{\mathrm{s}}^{(*)+}\mathrm{D}_{\mathrm{s}}^{(*)-}$ proposed in
Ref.~\cite{aleksan} has also been used by ALEPH.~\cite{aleph_dgs}

Methods based on double exponential lifetime fits to samples which contain a
mixture of CP eigenstates
%(\ie, inclusive, semileptonic, and $\Ds$-hadron),
have a quadratic sensitivity to $\dgs$, whereas methods based on isolating
a single CP eigenstate have a linear dependence on $\dgs$.
%($\phi\phi$, and J/$\psi\phi$), 
The branching ratio measurement mentioned above has also a linear dependence
on $\dgs$. The two last methods are therefore, in principle, more sensitive to $\dgs$, 
and are the only ones sensitive to the $\dgs$ sign. These methods, however, 
suffer from reduced statistics.
%the same applies to the branching ratio measurement mentioned
%above. The latter are therefore, in principle, more sensitive to $\dgs$, 
%and are the only sensitive to the $\dgs$ sign, but suffer from reduced statistics.

To obtain an improved limit on $\dgs$, the results based on proper time
distributions fits are obtained with an additional constraint on the allowed
range for $1/\Gamma_{\sq}$. The world average $\Bs$ lifetime is not used, as 
its meaning is not clear if $\dgs$ is non-zero. Instead, it is chosen
%(from theoretical motivations)
to constrain $1/\Gamma_{\sq}$ to the world average $\tau_{\Bd}$ lifetime:
$1/\Gamma_{\sq}\equiv1/\Gamma_{\dq}=(1.562\pm0.029)\ps$ %.\cite{hf_note}
(this assumption is motivated by theory).\cite{hf_note}

A description of all the analyses available, along with the results obtained
with each of them is found in Ref.~\cite{hf_note}

\begin{figure}
  \center
  \mbox{\epsfig{figure=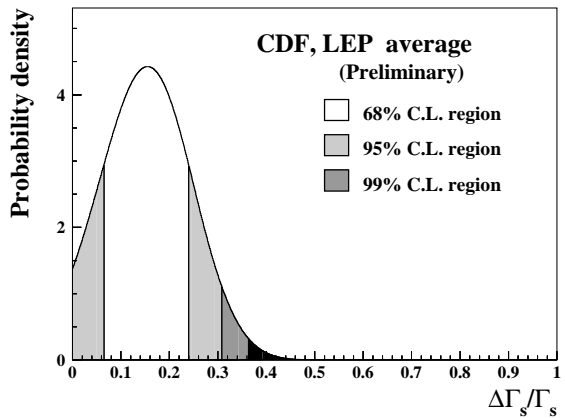,width=7.5cm}}
  \caption{Probability density function for $\dgs/\Gamma_{\sq}$ with the
    \mbox{$1/\Gamma_{\sq}\equiv\tau_{\Bd}$} constraint. The three shaded regions show
    the limits at $68\%$, $95\%$, and $99\%$ $\CL$~respectively.}
  \label{fig:dgs_comb}
\end{figure}
The world combined probability density function for $\dgs/\Gamma_{\sq}$ is
shown in Fig.~\ref{fig:dgs_comb}.
From this distribution an experimental limit and first measurement of
$\dgs$ is obtained:
\bea
\nonumber
\dgs/\Gamma_{\sq}\,=\,0.16^{+0.08}_{-0.09} \\
\dgs/\Gamma_{\sq}\,<\,0.31~~{\mathrm{at~95\,\%\,\CL}}~,
\label{eq:dgs}
\eea
which is in good agreement with the theoretical prediction given
in Ref.~\cite{Hashimoto}:
\bea
\dgs/\Gamma_{\sq}\,=\,0.097\,^{+0.038}_{-0.050}~.
\eea

The ratio $\dgs/\dms$ is calculated on the lattice, the value
obtained in Ref.\cite{Hashimoto} is
\mbox{$\dgs/\dms=(3.5\,^{+0.94}_{-1.55})\times 10^{-3}$},
%dgs/gs = 0.097 (+0.014-0.035) +/- 0.025 +/- 0.020 +/- 0.016
%dgs/dms = 3.5 (+-/4-1.3) +/- 0.6 +/- 0.6
which combined with the above $\dgs$ experimental results provides a mild constraint
on the allowed range for $\dms$:
\bea
\nonumber
%\dms \,=\,16\pm10\,\psin \\
%\dms\,<36\,\psin~~{\mathrm{at~95\%\,\CL}}~.
\dms \,=\,29\,^{+16}_{-21}\,\psin ~.
%\dms\,<66\,\psin~~{\mathrm{at~95\%\,\CL}}~.
\eea

\boldmath
\section{The $\Bs$ Oscillations}\unboldmath
\label{sec:dms}

\subsection{The Amplitude Method}
\label{sec:amp}
The $\Bs$ oscillation analyses performed so far are not able to resolve
the fast oscillation frequency $\dms$ and can only exclude a certain range
of frequencies. The combination of such excluded ranges is not straightforward, and
a specific method, the {\sl Amplitude Method}, has been developed for this
purpose.\cite{moser,ours} An amplitude $\amp$ is introduced in front of
the oscillating term of the probability density function for unmixed and
mixed $\Bs$ mesons:
\begin{eqnarray}
  \label{eq:ampl}
  \boldmath
  \nonumber
  p.d.f.^{\mathrm{u,m}}(t)=\frac{\Gamma_{\mathrm {s}} 
    e^{-\Gamma_{\mathrm {s}} t}}{2}\left [ 1\pm \cos (\dms t) \right ] ~\Rightarrow \\
  p.d.f.^{\mathrm{u,m}}(t)=\frac{\Gamma_{\mathrm {s}} 
    e^{-\Gamma_{\mathrm {s}} t}}{2}\left [ 1\pm{\bf \amp}\cos (\omega t) \right ]~.
  \unboldmath
\end{eqnarray}
The amplitude $\amp$ is measured for any value of the test frequency
$\omega$, with its uncertainty $\siga$. The value \mbox{$\amp=0$} is expected far below
the true oscillation frequency, and \mbox{$\amp=1$} at \mbox{$\omega=\dms$}. Frequencies for
which \mbox{$\amp+1.645\siga\leq1$} are excluded at $95\%\,\CL$ The expected limit for
a given analysis (also known as sensitivity) is defined as the frequency for
which $1.645\siga=1$.

\begin{figure}
\center
\mbox{\epsfig{figure=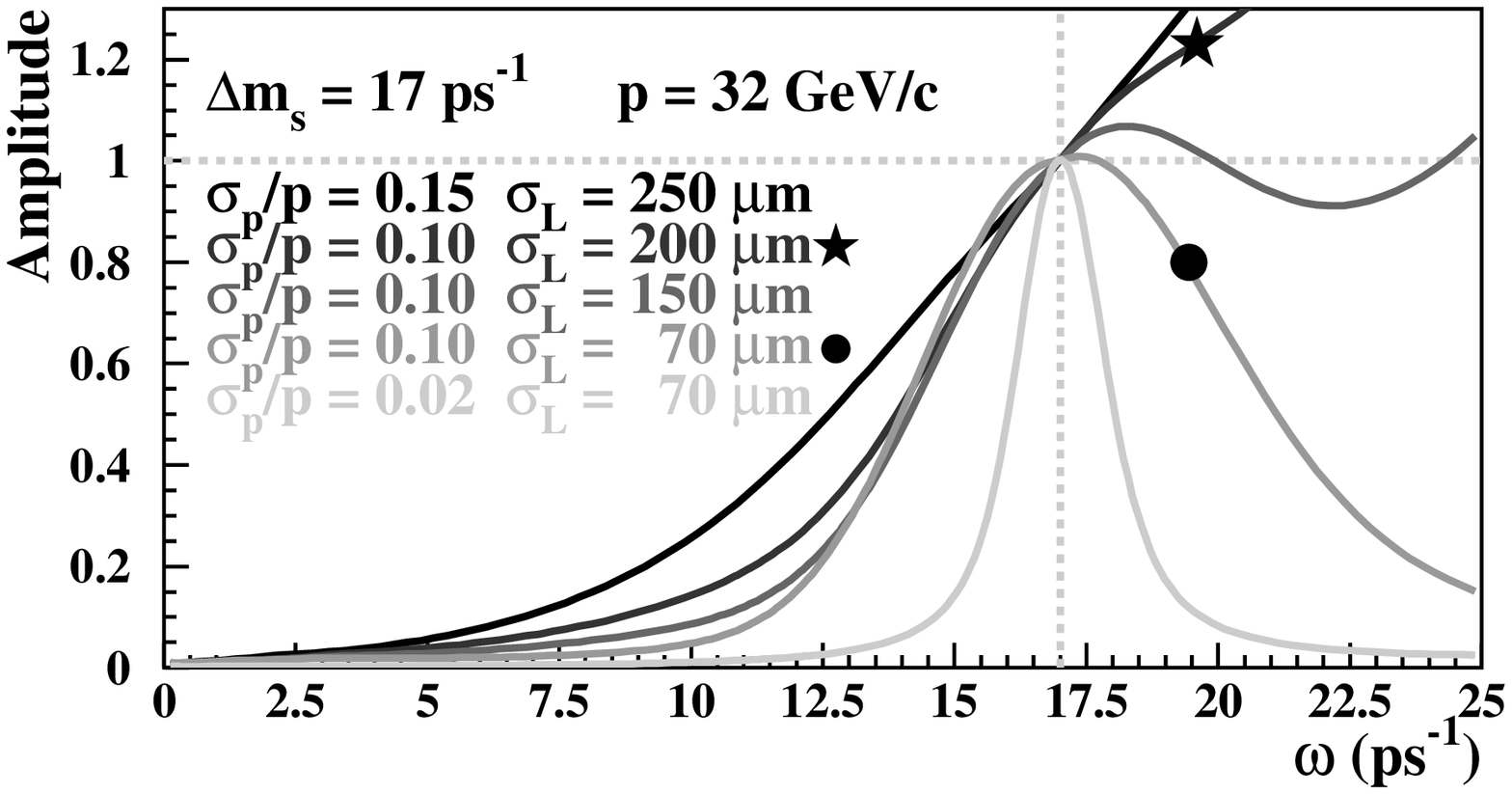,width=8.5cm}}
\caption{Expected amplitude shape for \mbox{$\dms=17\,\psin$}, \mbox{$p_{\B}=32\,\gev$}
and different resolution values.}
\label{fig:exp_amp}
\end{figure}

The expected amplitude shape for $\dms=17\,\psin$, $p_{\B}=32\,\gev$, and different 
resolution values are obtained from analytical calculations;\,\cite{ours} these
curves are shown in Fig.~\ref{fig:exp_amp}.
The typical resolutions achieved at LEP correspond to the  curve marked with a 
%star and those at SLD to 
star, while SLD is represented by 
the curve marked with a circle.

\subsection{Analysis methods}

The first step for a $\Bs$ meson oscillation analysis is the selection of final
states suitable for the study. The choice of the selection criterion determines
also the strategy for the tagging of the meson flavour at decay time. Then, the
flavour at production time is estimated, to give the global mistag 
probability $\eta$.
%The average $\Bs$ fraction is roughly $10\%$, 
The production rate of $\Bs$ mesons in the fragmentation of high-energy b quarks
is about $10\%$.
%Some analyses apply several techniques
%to increase this value and enrich the $\Bs$ content of the final sample.
In some analyses the selection of the final state chosen yeilds a higher
$\Bs$ content. In inclusive analyses several techniques are used to 
effectively increase the $\Bs$ content of the sample selected.

Finally, the proper time  is reconstructed for each meson candidate, and the
oscillation is studied by means of a likelihood fit to the distributions of the
decays tagged as mixed or unmixed.

The statistical power of a $\Bs$ analysis is described by the uncertainty on the measured
amplitude as a function of the test frequency $\omega$; 
it can be written as:
%it depends on the analysis items described above as:
\bea
S\,=\,\sigma_{\amp}^{-1}\propto\,\sqrt{N}\fs(1-2\eta){\mathcal{F}}(\omega,\sigma_t)~,
\eea
where $N$ is the total number of events, $\fs$ is the effective fraction of $\Bs$,
and ${\mathcal{F}}$ is a function increasing fast with $\omega$. The proper time
resolution $\sigma_t$ is expressed as a function of the decay length and momentum
resolutions, $\sigma_l$, $\sigma_p/p$, as:
\bea
\sigma_t\,=\,\frac{m}{p}\sigma_l\oplus\frac{\sigma_p}{p}t~.
\eea
The statistical power %significance 
of an analysis at high frequency (where the actual interest is)
is mainly determined %dominated 
by the proper time resolution; while number of events, tagging and
$\Bs$ enrichment contribute 
an overall factor to $S$, independent of frequency.
%to the statistical power at low frequencies.

The analysis methods used so far are explained below.  
 Up-to-date references for all
the $\Bs$ oscillation analyses are available from:
{\tt http://lepbosc.web.cern.ch/LEPBOSC/}.

\boldmath
\subsubsection{Fully reconstructed $\Bs$ candidates}\unboldmath

\noindent Specific hadronic $\Bs$ decays to flavour eigenstates such as 
$\Bs\rightarrow\Ds\pi$ and 
$\Bs\rightarrow\Ds {\mathrm{a}_1}$ can be fully reconstructed. No more than a 
few dozen such events have been reconstructed but their proper time resolution is 
so good that such analyses, in spite of their low sensitivity individually, contribute
at high frequency.
Only DELPHI~\cite{del_excl} and ALEPH~\cite{ds_lep_a} have released results with this method.

\boldmath
\subsubsection{Semi-exclusive samples: $\Ds\ell$, $\Ds$-hadron, $\phi\ell$}\unboldmath

\noindent Specific $\Ds$ decay channels are reconstructed and vertexed with
a lepton (or hadron) to form the $\Bs$ candidates. 
Statistics are relatively limited, typically a few hundred events for
a LEP experiment, to be compared to several ten thousand events in the
inclusive lepton analyses explained below. On the other hand,
the full reconstruction
of the $\Ds$ improves significantly the decay length resolution, and the $\Bs$ 
purity can be as high as $40\%$. The lepton sign is used for final state tagging.
Results from ALEPH, DELPHI, and OPAL are available.~\cite{ds_lep_a,ds_lep_d,ds_lep_o}
At SLD, due to their overall lower statistics (with respect to a LEP experiment),
a $\Ds\ell$ analysis is not competitive, instead a $\Ds$-hadron analysis, where
the $\Ds$ candidates are vertex with a charged hadron, is performed.\cite{sld_ds}
Finally, CDF and DELPHI have analyses~\cite{ds_lep_d,phi_lep} where
$\phi$-lepton correlations are exploited.

\subsubsection{Inclusive semileptonic samples}
\noindent High-$p_T$ leptons are searched for and vertexed together with an inclusively 
reconstructed charmed particle. 
This method profits from high statistics, reasonable resolution (depending
\begin{figure}[bh]
  \center
%  \mbox{\epsfig{figure=comp_best_bw_2.eps,width=7.2cm}}
  \mbox{\epsfig{figure=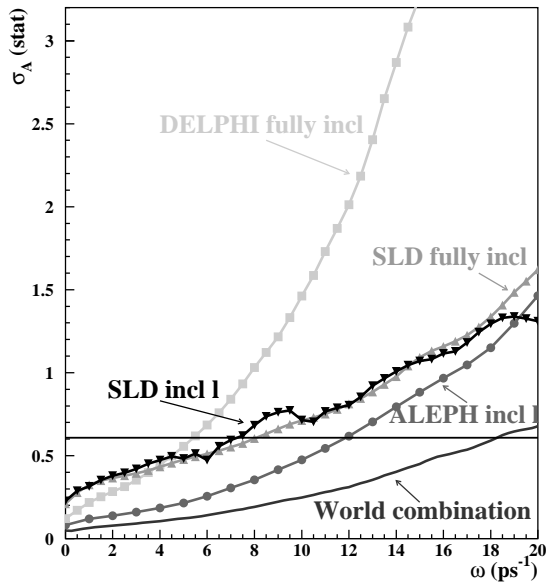,width=7.2cm}}
  \caption{Comparison of the statistical power of the best inclusive lepton 
    and fully inclusive analyses at LEP and SLD. }
  \label{fig:lep_sld}
\end{figure}
on the charm selection) but low $\Bs$ purity, at 
the level of $\fs\approx10\%$. The final state tag
is given by the lepton sign.
Results from ALEPH, DELPHI, OPAL, and SLD are available with this 
method.~\cite{a_inc_lep,del_osaka,op_incl,sld_incl}

Inclusive lepton and $\Ds$-lepton are the most sensitive analysis methods at LEP.

\boldmath
\subsubsection{Fully  inclusive samples}\unboldmath

\noindent The b hadron is reconstructed inclusively. The final state tag is
computed with inclusive charge estimators from the decay hemisphere. 
In particular, in SLD, thanks to their excellent resolution, 
the secondary and the tertiary vertices are reconstructed and the
charge flow between the two  is used to separate $\bz$ from $\bzb$.
%Decay vertices are reconstructed inclusively. In each hemisphere a secondary 
%(B) and a tertiary (D) vertex are searched for. 
%The final states tag is determined using a {\it dipole} technique: the charge flow
%between the two vertices is used to separate $\bz$ from $\bzb$.
This method benefits from the highest statistics. % with reasonable resolution.
Only SLD and DELPHI have released results with such a method.~\cite{sld_incl,del_osaka}

Inclusive lepton and fully inclusive analyses are the most sensitive at SLD.

\subsection{Analysis comparison}
The most competitive results to date on $\Bs$ oscillations are obtained by the LEP 
experiments and SLD. The differences between the two environments %, LEP and SLD,
are  illustrated in Fig.~\ref{fig:lep_sld}.
%\begin{figure}
%  \center
%  \mbox{\epsfig{figure=comp_best_bw_2.eps,width=7.2cm}}
%  \caption{Comparison of the statistical power of the best inclusive lepton analysis
%    at LEP and SLD. The same is shown for the fully inclusive analyses.
%    }
%  \label{fig:lep_sld}
%\end{figure}
The statistical error on the measured amplitude is plotted as a function of the
test frequency $\omega$. The best inclusive lepton analysis from LEP is equivalent
to the corresponding SLD one at high frequency. However, in the case of a fully inclusive
analysis, the SLD experiment has no rival.

\begin{figure}[bh]
  \center
%  \mbox{\epsfig{figure=compl_bw_2.eps,width=7.2cm}}
  \mbox{\epsfig{figure=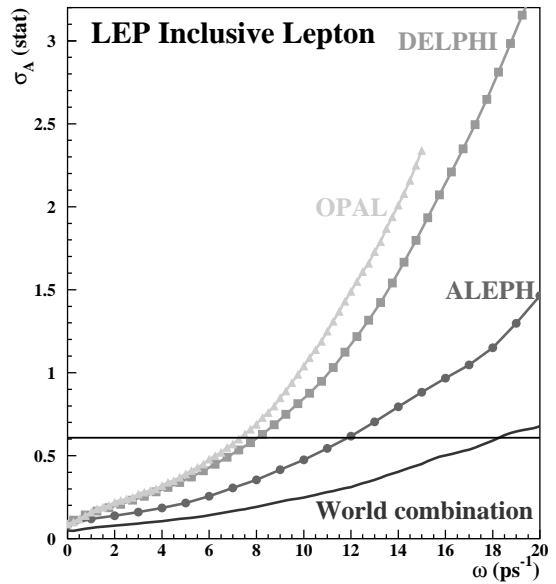,width=7.2cm}}
  \caption{Inclusive lepton analyses comparison between LEP experiments.
    }
  \label{fig:incl_lepton_lep}
\end{figure}

The inclusive lepton analyses from  ALEPH, DELPHI and OPAL are compared in
Fig.~\ref{fig:incl_lepton_lep}.
The statistical power of the three analyses is similar at low frequency, but at high
frequency the ALEPH 
analysis has a significantly better performance. The bulk of the difference
is not related with the detector performance, but is rather due to the
analysis technique: specifically, ALEPH achieves a higher statistical
power also thanks to a more detailed event-by-event estimation of the
uncertainty on the reconstructed $\Bs$ decay length and momentum.
%results are more competitive. The differences are mostly not detector
%related, but rather due to a different analysis approach: in ALEPH a careful event-by-event
%treatment of the vertexing is performed.

\subsection{Present Results}
\begin{figure}
\center
\mbox{\epsfig{figure=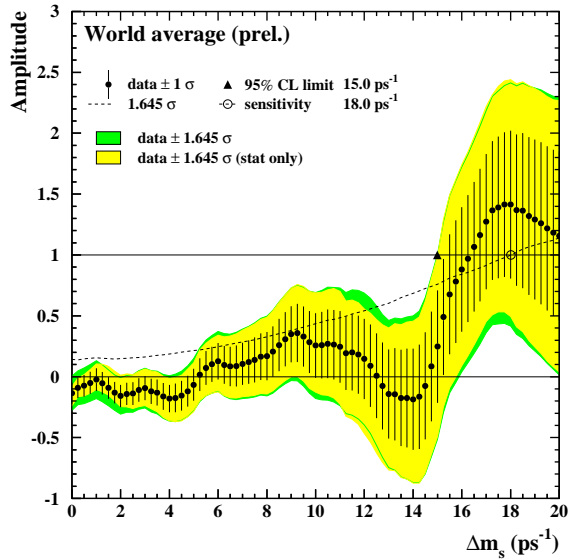,width=7.5cm}}
\caption{World average amplitude spectrum as for summer 2000.}
\label{fig:w_ave}
\end{figure}

All available $\Bs$ oscillations analyses  are combined to obtain the
amplitude spectrum  shown in Fig.~\ref{fig:w_ave}.
In one year, from summer 1999 to summer 2000, an overall sizable improvement is observed:
the expected limit has gone up by $3.5\psin$, or in other words, the statistical error on the
amplitude at high frequency has been reduced by a factor of 2. 
Most of this improvement is given by the new results from ALEPH and 
SLD.\cite{a_inc_lep,sld_incl,sld_ds}
The limit set at $95\%\,\CL$, however,
has not followed this increase due to the presence of a $2.3\sigma$ deviation from $\amp=0$ 
around $\dms=17\psin$. The probability that this deviation be due to a statistical fluctuation
is estimated to be at the $3\%$ level with gedanken experiments.

\subsection{Perspectives for the near future}

By next summer final results from present analyses at LEP (ALEPH and DELPHI) and
SLD are expected. The combined sensitivity could increase by few inverse picoseconds and
therefore the significance of the signal hint observed  could also increase.

In principle more $\Bs$ oscillations studies could be performed with the LEP data, but
might remain undone.

\subsection{Conclusions}

The LEP, SLD and CDF collaborations have greatly contributed to $\Bs$ physics. A
summary of our present knowledge is shown in Table~\ref{tab:results}, the value of the
measured $\Bs$ meson mass and lifetime are included for completeness.
\begin{table}
\begin{center}
\caption{Summary of $\Bs$ experimental knowledge.}\label{tab:results}
\vspace{0.2cm}
\begin{tabular}{|l|} 
\hline 
$m_{\Bs}\,\,=\,\,5369.6\,\pm\,2.4\,\mev$ \\
$\tau_{\Bs}\,\,=\,\,1.464\,\pm\,0.057\,\ps$ \\
$\dgs/\Gamma_{\sq}\,\,=\,\,0.16\,^{+0.08}_{-0.09}$\\
$\dms \,\,>\,\,15\,\psin~{\mathrm{at}}~95\%\,\CL$\\
\hline
\end{tabular}
\end{center}
\end{table}

%The $\Bs$ oscillations excitement is still on the LEP and SLD side, but
%soon CDF and D0 will take over and, if not done before, measure $\dms$.
Efforts to resolve $\Bs$ oscillations are still ongoing at LEP and SLD.
Soon CDF and D0 will take over and $\dms$ will probably be measured
(if not before) from the data collected in the Tevatron RunII

\section*{Acknowledgements}
It is a pleasure to thank the organizers of the BCP4 Conference for their
wonderful hospitality in Ise-Shima. I wish to thank D.Abbaneo and R.Forty for their
careful reading of this manuscript. 
This work was supported by the Commission of the European Communities, contract
ERBFMBICT982894.

\end{document}